# ACDMCP: An Adaptive and Completely Distributed Multi-hop Clustering Protocol for Wireless Sensor Networks

Khalid Nawaz and Alejandro P. Buchmann

Databases and Distributed Systems Group, Department of Computer Science,
Technische Universität, Darmstadt, Germany
khalid@dvs.tu-darmstadt.de
buchmann@dvs.tu-darmstadt.de

## ABSTRACT

*Clustering is a very popular network structuring technique which mainly addresses the issue of scalability in large scale Wireless Sensor Networks. Additionally, it has been shown to improve the energy efficiency and prolong the life of the network. The suggested protocols mostly base their clustering criteria on some grouping attribute(s) of the nodes. One important attribute that is largely ignored by most of the existing multi-hop clustering protocols is the reliability of the communication links between the nodes. In this paper, we suggest an adaptive and completely distributed multi-hop clustering protocol that incorporates different notions of reliability of the communication links, among other things, into a composite metric and uses it in all phases of the clustering process. The joining criteria for the nodes, which lie at one hop from the elected cluster heads, to a particular cluster not only consider the reliability of their communication link with their cluster head but also other important attributes. The nodes that lie outside the communication range of cluster heads become cluster members transitively through existing cluster members utilizing the end-to-end notion of link reliability, between the nodes and the cluster heads, along with other important attributes. Similarly, inter-cluster communication paths are selected using a set of criteria that includes the end-to-end communication link reliability with the sink node along with other important node and network attributes. We believe that incorporating link reliability in all phases of clustering process results in an efficient multi-hop communication hierarchy that has the potential of bringing down the total communication costs in the network.*

## 1. INTRODUCTION

Wireless Sensor Networks (WSN) are composed of tiny Micro-electromechanical Sensing (MEMS) devices that have a potential use in many different application scenarios. They are normally used for collecting and processing environmental data, and detecting and reporting events of interest to some base station which normally is more resource rich than these sensing devices. The number of these tiny sensing devices used in a given application could vary from tens of devices to possibly hundreds or even more. This important aspect of scale raises complex issues regarding efficient use of the network and nodes' resources during the operation of the network. This issue is generally addressed by grouping the nodes into clusters, thus defining energy efficient communication paths both within and among the clusters. Additionally, improving energy efficiency, which consequently results in prolonging the life of the network, has been touted as another important goal of clustering.

Some other goals that are attributed to clustering include bandwidth reuse in the network, efficient data gathering and aggregation, target tracking, and supporting hierarchical routing techniques. Moreover, in some middleware approaches like [1, 20] clustering not only helps in solving the scalability issue but it also helps in increasing the event reporting reliability. Additionally, clustering has also been proposed for solving security issues in mobile adhoc networks as suggested in [21].

The clustering protocols that have been suggested so far could be grouped into two broad categories based upon cluster formation criteria and the parameters used for Cluster Head (CH) election [2]. These categories are: probability based (random or hybrid) and non-probability based clustering protocols. In the probability based protocols, each sensor node is assigned a probability of becoming a CH, either randomly or based upon some attributes of the node, and higher probability nodes constitute the initial set of CHs. These protocols normally are iterative, raising probabilities of a node to become a CH in each round, thus converging, in some specific number of iterations, to a final set of CHs. Some protocols in this category randomly elect CHs [3] without paying attention to either residual energy or any other relevant grouping attributes of the nodes. There are, however, some probability based protocols [4] that incorporate residual energy as a primary cluster head election parameter and some secondary parameter like node degree to supplement the cluster head election process.

In the second (non-probabilistic) category of clustering protocols, some specific criteria like node identifiers, connectivity or node degree is used to elect CHs. In both categories, once the CHs have been elected, the rest of the nodes use some criterion, like proximity to the CHs or the degree of the CHs, to join them to form clusters. However, one very important aspect that has largely been ignored by many of the existing multi-hop clustering protocols is the reliability of the communication links between the nodes.

One of the fundamental goals of clustering techniques is to generate energy efficient communication hierarchies that define communication paths for routing data through the network. If a clustering protocol generates communication paths, ignoring the link reliability between the nodes, then the resulting communication hierarchy will, most probably, fail to achieve its most fundamental goal of being energy efficient due to higher message loss on unreliable paths. Most of the existing clustering techniques try to define clusters such that the nodes that form a cluster are physically close to each other. They implicitly believe that it is the node proximity that is the decisive factor in determining energy consumption in WSNs. However, it is not only the physical distance that matters when two nodes communicate with each other but there are a multitude of other factors that could influence the communication. For instance, two nodes that are in close proximity to each other might very well have a very poor communication link between them due to multipath interference or simply because they are located in that part of the network where node density is high, thus resulting in a high channel contention. Therefore, it is very naive to only consider the physical distance between nodes while defining a clustering communication hierarchy in WSNs.

Having said that, there are some single hop clustering protocols that incorporate communication reliability between nodes while choosing cluster heads [5, 18]. These protocols, however, are single-hop clustering protocols and their consideration of link reliability is mostly confined to just a subset of the clustering process. ACDMCP, on the other hand, makes link reliability an integral part of all phases of the clustering process, namely, CH election, cluster formation, and inter-cluster communication. On top of that, ACDMCP offers multi-hop clusters and the robust incorporation of link reliability in all phases of the clustering process ensures that the communication paths that are generated by the protocol in the network can reliably transport data.

ACDMCP belongs to the non-probabilistic category of clustering protocols, since no a priori probabilities are assigned to the nodes. Instead each node determines its Cluster Head Competence Value (*CHCV*), which is a composite weighted metric, similar in the weighting aspect to the one suggested in WCA [7]. This metric incorporates some important node level as well as network level attributes like the strength of the node's communication links in its neighborhood to determine its suitability in assuming the role of a CH. We show that such an approach results in a reliable communication hierarchy that is energy efficient and prolongs the life of the network. Each node, while joining a cluster, utilizes the *CHCV* metric but with a different notion of the link reliability than the one used while electing CHs. In case of a tie, the node successively compares the constituent attributes of the *CHCV* metric for choosing the best offer. If everything turns out to be equal, then the node breaks the tie using Node Ids. Transitive cluster membership, through existing cluster members, also involves utilizing *CHCV* metric but with an end-to-end reliability value replacing the out link reliability parameter that is used for deciding direct cluster membership.

## 1.1. Our Contribution and Paper Structure

This paper makes the following contributions:

1. It suggests a multi-hop clustering protocol that makes link reliability an integral part of all phases of the clustering process, thus increasing the chances of generating a communication hierarchy that offers more reliable communication paths for data transmission through the network.

2. The suggested protocol, apart from making communication reliability an integral part of the clustering process, also incorporates residual energy of the nodes throughout the clustering process, thus ensuring delegating more responsibilities in the hierarchy to the higher energy nodes.

3. The suggested multi-hop clustering mechanism makes it possible to share the cluster management load for the k-hop Cluster Members (also called transitive CMs) with the transitive CHs diverging from many existing approaches which put all cluster management load on the CHs only.

4. The adaptive nature of the suggested protocol allows nodes to switch clusters, if and when they receive a better offer. Clusters also evolve over time, since with each new round of clustering, nodes have link statistics collected over a longer period of time which enable them to make more informed decisions in subsequent rounds of clustering.

The rest of the paper is organized as follows. Section 2 presents some of the related approaches that have been suggested for clustering WSNs and/or Mobile Ad hoc NETworks (MANETS). It is followed by Section 3 which describes the system model used in the protocol including some important relevant definitions. Section 4 describes the cost metric used in the protocol. Section 5 describes the clustering process in detail along with the assumptions made and the requirements set for the protocol. Next there is a section on evaluation of the protocol, Section 6, which describes the simulation model and the results obtained by running the protocol in simulation. In Section 7 some of the applications of clustering in WSNs have been briefly mentioned. Finally in Section 8, we present some of the conclusions that we draw from the results.

## 2. RELATED WORK

There are quite a few clustering protocols that have been suggested for MANETS as well as for WSNs. Protocols like LCA [8] and WCA [7] which have been suggested for MANETS have limited applicability in WSNs due to their focus on mobility of the nodes than on other attributes which are more relevant in WSNs. For WSNs, one of the earliest clustering protocols is LEACH [3]. It belongs to the probabilistic category of clustering protocols and is distributed in nature. However, it assumes that each node is in one hop communication range of the sink node, an assumption that reduces its applicability to large scale WSNs. Additionally, in LEACH each node has an equal probability of becoming a CH. This is done in order to distribute the load of being a CH among all nodes by repeating the clustering process at fixed intervals. However, it is very likely that a node which is running low on battery could become a CH, thus increasing the chances of node failure and information loss in the network. Moreover, since it generates single hop clusters, the chances of having orphan nodes, which don't have a CH in range, are also there.

Another well known single hop clustering protocol is HEED (Hybrid Energy-Efficient Distributed Clustering) [4]. It uses different radio transmission power levels for intra-cluster and for inter-cluster communication, thus allowing multi-hop communication among the CHs to transport data to the sink node. Its consideration of residual energy of the nodes in the CH election phase, as opposed to LEACH, ensures that only high energy nodes are chosen as CHs. However, the nodes which lie in communication range of multiple CHs, don't consider residual energy or any reliability oriented attribute to join a cluster. Instead, they consider a secondary parameter, node degree, to make their cluster joining decision. Additionally, for inter-cluster communication, no specific node or network level attributes are considered. Though HEED improves on some of the limitations of LEACH protocol, it has its shortcomings in the cluster joining and inter-cluster communication phases.

Energy-Efficient Hierarchical Clustering (EEHC) [9] is a clustering protocol that takes into account the energy heterogeneity of the sensor nodes in the network. It divides nodes into three categories based upon their residual energy, namely, super, advanced, and normal nodes, with super and advanced nodes having higher energy levels than the normal nodes. Using this heterogeneity in energy levels of the nodes, authors present a mathematical model to assign weighted probabilities to nodes for the cluster head election phase. Except for considering heterogeneous energy levels of nodes, the rest of the protocol is very similar to LEACH. ACDMCP also makes no prior assumption to the homogeneity of the energy levels of the nodes.

Distributed Weight-Based Energy-Efficient Hierarchical Clustering (DWEHC) [19] is a multi-hop weight based clustering protocol that tries to minimize energy usage by allowing nodes choose either a first-level membership or a second-level membership depending upon their distance from each other. The basic assumption of the protocol revolves around the same idea that the energy consumption is a function of the distance between the nodes. It also doesn't consider any link quality measure in the clustering process.

There are some hierarchical routing approaches like PEGASIS [10], TEEN [11], APTEEN [12] which have been suggested for energy conservation in WSNs. The basic PEGASIS protocol, organizes nodes in a chain structure rather than organizing them in clusters. It assumes that every node has a global knowledge of the position of all nodes in the network, an assumption that limits its applicability in large scale WSNs. The nodes' adaptive transmission power control is used to communicate within the chain as well as to the base station. A chain leader is elected in each round to which all nodes send their data in a multi-hop fashion and after aggregating it, the chain leader transmits it to the base station in a single hop. However, its assumptions like having a global knowledge of the positions of all nodes, no consideration of energy in choosing the chain leader and that each node can directly communicate with the base station limits its applicability to any reasonably sized network. TEEN uses LEACH protocol to build clusters of homogeneous nodes with the same initial energy reserves. It is especially suited to reactive networks, which respond to the changes in the parameter of interest immediately as opposed to the more passive proactive networks that gather data. It defines soft and hard thresholds for the nodes to report data to their corresponding CH, which aggregates and forwards the data to the base station. Again the emphasis here is on building clusters without considering any reliability oriented parameter.

APTEEN is an extension of the TEEN protocol and uses LEACH for clustering the network. The only improvement from TEEN is that the network is assumed to have both reactive and proactive features. Thus the protocol offers mechanisms to make the network report periodic data as well as near real-time reactive reporting about critical events.

## 3. SYSTEM MODEL AND IMPORTANT DEFINITIONS

We model the WSN as a directed graph $G(V, E)$ with the set of vertices ($V$) representing the set of nodes in the network and the set of edges ($E$) representing communication links between the nodes. The communication links are symmetric in the sense that if a node ($V_1$) can communicate with a node ($V_2$), then the node ($V_2$) can also communicate with the node ($V_1$). However, the links in terms of communication reliability are not necessarily symmetric.

Communication link reliability $\left(\lambda_{\varepsilon_{ij}}\right)$ represents the probability of a successful packet transmission from a node ($V_i$) to a node ($V_j$) having a communication link $\varepsilon_{ij}$, where nodes $V_i$ and $V_j$ lie in each others' communication range. For instance, if the communication link reliability between two nodes ($V_1$ and $V_2$) in the direction of $V_1 \rightarrow V_2$ is $\left(\lambda_{\varepsilon_{12}}\right)$, it might not be the same in the opposite direction from $V_2 \rightarrow V_1$, $\left(\lambda_{\varepsilon_{21}}\right)$ i.e. $\lambda_{\varepsilon_{12}} \neq \lambda_{\varepsilon_{21}}$.

Therefore, all edges that are incident upon a vertex (node) represent in-Link Reliability (in-LR) of the node with its one hop neighbors. Similarly, all edges that emanate from a vertex (node) represent out-Link Reliability (out-LR) of the node with its one hop neighbors.

In the multi-hop case, the directed path $\vec{m}$ between two nodes $V_i$ and $V_m$ of length d-hops is represented by $\vec{m}\left(\varepsilon_{ij}, \varepsilon_{jk}, \varepsilon_{kl}, \varepsilon_{lm}\right)$, where the directed edges from $V_i$ to $V_m$ are given in their order of appearance starting from the node $V_i$. Since link reliability is a multiplicative metric, on a multi-hop directed path $\vec{m}$, it is the product of the link reliabilities of the constituent links of the multi-hop directed path $\vec{m}$. Consequently, end-to-end communication Link Reliability $ELR(V_i, V_m)^d$ between two nodes $V_i$ and $V_m$ that lie d-hops from each other is the product of the link reliabilities of all the communication links that make up the d-hop path between $V_i$ and $V_m$.

$$ELR(V_i, V_m)^d = \prod_{\varepsilon_m = 1}^{d} \lambda_{\varepsilon_{\vec{m}}} \qquad (1)$$

The neighborhood set of a node ($V_i$), represented as N($V_i$), is composed of the nodes that the node $V_i$ can directly communicate with in a single hop at some specified transmission power level $(P_{T_x})$. The neighborhood sets of neighboring nodes overlap with each other. However, two neighborhood sets must have at least one unique member to make them unique, i.e. $N(V_i) = N(V_j)$, iff $V_i = V_j$. The size of the neighborhood set of a node $V_i$ is called the degree of the node $D_{V_i}$. The set of all communication links for a node $V_i$ with each of the nodes in its neighborhood set is represented as $E(V_i)$. The values of the in/out-link reliabilities on each of these links can vary depending upon factors like transmission power, presence of obstacles, multipath interference, and presence of other devices communicating with each other at frequencies in the ISM band. The average of these values over all the links in $E(V_i)$ for a node $V_i$ is termed as the Mean in/out-Link Reliability $MLR_{in|out}(V_i)$.

$$MLR_{in|out}(V_i) = \frac{\sum_{j=1}^{k} \lambda_{\varepsilon_{ij}}}{k} \qquad (2)$$

Where, k is the size of the neighborhood set of the node $V_i$.

A Cluster Head (CH) is a node $V_i$, such that it has the highest $CHCV_{MLR_{in}(V_i)}$ (see Section 4) in its neighborhood set. A cluster is composed of any subset ($C$) of the set of nodes ($V$), such that all elements of the subset ($C$) are in either direct communication range of the CH ($V_i$) or are transitively reachable through some member of C, which we call Transitive Cluster Head (TCH). A Cluster Member (CM) is a node $V_i$ such that it can reach the CH in a single hop and that its $CHCV_{LR_{out}}$ is better than its $CHCV_{LR_{out}}$ with all other CHs that are reachable in a single hop from it. If $CHCV_{LR_{out}}$ of a node turns out to be equal for all the CHs that are at 1-hop from it, then the node successively compares the components of the $CHCV_{LR_{out}}$ of the CHs to choose the best one to join. A k-level CM of a cluster is a node $V_i$ that joins the cluster through some existing CM or some (k-1)-level CM, which becomes its TCH. The k-level CMs also compare the $CHCV_{ELR_{out}}$ of all 1-hop neighbors which are already CMs or (k-1)-level CMs and which have made a cluster joining offer.

## 4. ACDMCP Cost Metric

ACDMCP uses a weighted composite metric (*CHCV*) that incorporates important node and network oriented attributes. The constituent attributes of this metric are converted to indices, whose values vary between 0 and 1, using schemes that ensure the desired contribution and impact of the attributes in the metric.

### 4.1. Residual Energy Index (REI)

In almost all clustering approaches, the nodes that form a cluster report their data to the elected CHs, which normally aggregate and forward the data to the base station either directly in a single or multiple hops. The role of the CH, thus, has some additional responsibilities which put higher demands on its already constrained energy reserve. In some approaches, CHs are assumed to have better resources, e.g. energy, than the normal nodes in the network. However, many clustering approaches consider all nodes to have homogeneous resources. Therefore, apart from few earlier clustering approaches, many recent approaches consider residual energy of the nodes while choosing CHs. They, however, mostly try to relate power consumption with the distance between the nodes and not with the communication link reliability.

ACDMCP also considers residual energy ($E_{re}$) of the nodes. However, unlike some other approaches, it doesn't make any prior assumptions about the energy homogeneity of all the nodes. It also takes a broader view of power consumption and doesn't confine it to just the distance between the communicating nodes. It also defines a threshold value for the energy ($E_{th}$) of the nodes which marks the bare minimum energy level a node should possess, if it is to take part in the CH election process. This is a design parameter of the protocol and can be tuned appropriately to suit the application needs and expected load on the CHs. As soon as the $E_{re}$ of a CH falls below the $E_{th}$, it gives up its CH status and calls for reclustering the network.

**Algorithm 1** DETERMINE REI
1: **if** ($E_{re} > E_{th}$) **then**
2:     **if** (($E_{re} - E_{th}$)/$E_{th}$) > 1.0 **then**
3:         REI = 1.0
4:     **else**
5:         REI = (($E_{re} - E_{th}$)/$E_{th}$)
6: **else**
7:     REI = 0.001

In each round of clustering, the $E_{th}$ value is reduced by a specific percent of its value in the last round. This is necessary to ensure that there are nodes that have higher $E_{re}$ than the $E_{th}$, which could contest the CH election. It could, however, be argued that a CH whose $E_{re}$ falls below the $E_{th}$, might become a CH again in the subsequent round, if $E_{th}$ is dropped by a certain amount in each round of clustering. It could seldom happen though, since CHs consume more energy in carrying out additional duties. So, even if a CH whose $E_{re}$ has fallen below the $E_{th}$ and which has called for reclustering the network, subsequently has a higher $E_{re}$ than the $E_{th}$ after lowering the later, it might not necessarily be the best node in its neighborhood to assume the CH role again.

Each node is assigned an REI between 0 and 1, as is given in the Algorithm 1, where 1 is the best value. However, if a node's $E_{re}$ is below the $E_{th}$, it is assigned a minimum value of 0.001 to eliminate its chances of becoming a CH.

### 4.2. Node Degree Index (NDI)

Another aspect that we incorporate in the *CHCV* metric is the size of the node's neighborhood (node degree). Node degree gives an indication of the possible size of the 1-hop membership of a cluster, should the node become a CH. Therefore, in the CH election and cluster joining phases of the clustering process, node degree is used in the *CHCV* metric.

**Algorithm 2** DETERMINE NDI
1: **if** node degree = IDEG **then**
2:     NDI = 1.0
3: **else if** node degree > IDEG **then**
4:     NDI = IDEG/node degree
5: **else**
6:     NDI = node degree/IDEG

We use the notion of Ideal DEGree (*IDEG*), similar to WCA [7], which serves the purpose of giving more importance to the nodes that have the desired degree in becoming CHs. This could be used as a load balancing mechanism in the cluster joining phase. Similar to *REI*, each node computes Node Degree Index (*NDI*) (as depicted in Algorithm 2) whose value varies between 0 and 1, with 1 being the best value. This scheme ensures that the nodes with degrees equal to the *IDEG* get a maximum value of 1 and other nodes always get a value lower than 1. In the inter-cluster communication phase, however, we don't use node degree to compute *NDI*. There we use each CH's actual 1-hop membership count instead. This scheme could be exploited to our benefit in two different ways. Firstly, if CHs strive for a higher degree of aggregation in the network, they can choose those CHs, as their downstream neighbors, which have a higher 1-hop membership count. On the other hand, if the goal is to balance the load and consequently energy consumption, then CHs with a lower 1-hop membership count could be chosen as downstream neighbors because they have less cluster management load and would possibly have a higher $E_{re}$ too.

### 4.3. Link Reliability and hop-distance

The third component of the *CHCV* metric depends upon the node's strength of communication links with its 1-hop neighbors, i.e. link reliability. Two types of metrics could be used to assess reliability of a communication link, namely, hardware based and software based [6]. Examples of the hardware based metrics are Link Quality Indicator (LQI), Received Signal Strength Indicator (RSSI), and Signal to Noise Ratio (SNR). These metrics are easy to obtain directly from the radio hardware. However, they are calculated by the radio hardware only for the successfully received packets and that too on the basis of the first 8 symbols of the received packet. They have also been shown to be inadequate to properly represent the quality of a communication link [13].

Among the software based link reliability metrics, the most well known is the Packet Reception Ratio (PRR) and its derivatives. Therefore, we also use PRR in ACDMCP to quantitatively represent the reliability of a communication link.

The hop-distance, (represented by $\varsigma$ in Equation (5)) between either a k-level CM and its CH or a node and its downstream neighbor in the inter-cluster communication, can influence communication costs. Although the multiplicative nature of the end-to-end link reliability takes care of the hop-distance implicitly, it fails to distinguish between two multi-hop routes where link reliability has a maximum value of 1. We, therefore, incorporate hop-distance into the *CHCV* metric (Equation (5)) wherever multi-hop paths are involved. This enables the nodes, while evaluating either k-level cluster membership offers or potential downstream neighbors in the inter-cluster communication, to minimize their hop-distance and, thus, lower the communication cost.

#### 4.3.1. Link Reliability Variants

The CHCV metric slightly differs in each phase of the clustering process. In the CH election phase, Mean in-Link Reliability ($MLR_{in}$) is used to calculate it and is accordingly represented as $CHCV_{MLR_{in}}$.

$$CHCV_{MLR_{in}} = REI \times IF_{REI} + NDI \times IF_{NDI} + MLR_{in} \times IF_{MLR_{in}} \tag{3}$$

The inclusion of $MLR_{in}$ in Equation (3) guarantees that the nodes which are strongly connected with their 1-hop neighbors will have higher chances of getting elected as CHs. Please note that we include the in-link reliabilities to compute *MLR* because we assume that most of the information flow, in a clustered network, is in the downstream direction, i.e. from nodes to the CHs and then towards the sink. It, however, doesn't limit the application of our protocol, in any way, to the scenarios where information flows in both directions. In such cases, *MLR* is computed by considering both in and out link reliabilities of the edges that are incident upon or are emanating from a node.

In the cluster formation phase, each node, lying at 1-hop from elected CHs, uses $CHCV_{LR_{out}}$ (Equation (4)) which is obtained by replacing $MLR_{in}$ in Equation (3) with out-Link Reliability of the node with the elected CHs to evaluate their cluster joining offers.

$$CHCV_{LR_{out}} = REI \times IF_{REI} + NDI \times IF_{NDI} + LR_{out} \times IF_{LR_{out}} \quad (4)$$

The nodes that are not in direct communication range of any CH join clusters transitively through existing CMs or TCMs. Their cluster joining decision is also based upon a mix of attributes which is represented by Equation (5). In the same way, when the inter-cluster communication paths are formed, CHs use $CHCV_{ELR_{in|out}}$, which is obtained by replacing *MLR*$_{in}$ in Equation (3) with end-to-end out/in-Link Reliability $\left(ELR_{in|out}\right)$ of the CH with the sink node, to choose the best downstream neighbor.

$$CHCV_{ELR_{in|out}} = REI \times IF_{REI} + NDI \times IF_{NDI} + ELR_{in|out} \times IF_{ELR_{in|out}} + 1/\varsigma \times IF_{\varsigma\varsigma} \quad (5)$$

It is evident from the above three equations (3, 4 and 5) that ACDMCP incorporates relevant notions of link reliabilities in all phases of the clustering process. Additionally, it incorporates node's residual energy to make sure the nodes with higher energy get preference in all phases of the clustering process.

The impact of the constituent parameters of *CHCV* can be controlled by varying the values of the Impact Factors (IFs), which are similar to the weights assigned in clustering approach for MANETs [7], in the above equation. The value of each of these IFs varies between 0 and 1, and like any weighted average their sum equals 1. We, however, avoid assigning a value of 0, which would effectively remove the influence of that particular parameter from the metric. However, if a given application demands maximum focus on only one of these parameters, then an IF of 1 could be assigned to that parameter. The incorporation of IFs in *CHCV* adds flexibility and allows one to choose an appropriate mix which suits the given application scenario.

## 5. PROTOCOL DETAILS

Before describing the protocol details, it is important to lay down the basic set of requirements that the protocol should meet. In the next subsection, we list down some fundamental requirements that led to the design of ACDMCP.

### 5.1. Requirements on the Clustering Protocol

In our opinion any given clustering protocol should execute in a completely distributed fashion in the network to save the overhead of collecting global knowledge of the network at the sink. Another important consideration is to favor the high energy nodes and the nodes with better connectivity and communication reliability in becoming CHs. Therefore, the nodes that have more reliable communication links in their respective neighborhoods and those which also have higher residual energy should have higher chances of becoming CHs. It should also be tunable vis-à-vis the parameters that make up the metric used for choosing CHs, i.e. it should be possible to change the importance given to different parameters used for choosing CHs [7]. For instance, if the requirement is to give higher chances to the nodes with high energy in becoming CHs, then it should be possible to easily accomplish that.

Since one of the fundamental reasons of clustering the network is to solve the scalability problem, the protocol's overhead should be independent of the size of the network. Additionally, the protocol should not need extra knowledge like geographical locations, distance and direction of nodes as some of these requirements require extra hardware on the nodes that can increase their cost. One important feature that is desired in a clustering protocol is its adaptivity vis-à-vis communication reliability. Therefore, the protocol should progressively choose more reliable communication paths in each successive round of clustering. Last but not least, the protocol should use as few resources as possible to respect the resource scarce nature of WSN.

## 5.2. ACDMCP Grouping Attributes

ACDMCP incorporates different incarnations of important node and network attributes in different phases of clustering. These attributes incorporate different aspects of the requirements that we have listed above. These attributes are: Node's Residual Energy ($E_{re}$), Node Degree $\left(D_{V_i}\right)$, Link Reliability $\left(\lambda_{\varepsilon_{ij}}\right)$, Transmission Power $\left(P_{T_x}\right)$, Hop Distance $\left(\varsigma\right)$ and Node ID (NID).

In ACDMCP, each node in the network could be in any of the following states: Un-clustered (UC), Cluster Head Candidate (CHC), Direct Cluster Member (CM), Transitive Cluster Member (TCM), and Cluster Head (CH). There is a special state in the protocol which is called Single Node Cluster Head (SNCH) which is assumed by only those nodes whose $N_{V_i}$ is an empty set. This could happen only, for instance, for those nodes all of whose neighbors have died because of low battery. The role of Transitive CH (TCH) could be assumed by CMs or TCMs. If there are nodes which don't have any CH in their 1-hop neighborhood, then they choose, based upon $CHCV_{ELR_{out}}$ metric, any existing CM or TCM, which makes a transitive cluster membership offer, as their transitive CH. The state transition diagram of ACDMCP in Figure 1 shows different states that a node can go through during the execution of the protocol.

## 5.3. Cluster Head Election

At the beginning of the protocol each node is in the UC state. The first task is to determine the neighborhood set $\left(N_{V_i}\right)$ of each node as well as the initial link reliabilities within its neighborhood set. It is achieved by randomly broadcasting Neighbor Discovery Messages (NDMs) with one of the lower transmission power levels $\left(P_{T_x}\right)$ available to the node. The higher transmission power levels are allocated for inter-cluster communication as in some of the other clustering protocols [4].

During this initial phase of determining $N_{V_i}$, each node sends "n" such broadcasts where "n" is a positive number whose value could be chosen depending upon the degree of certainty required for determining the link reliability. These repeated random broadcasts serve two purposes. For one, they help in determining accurate $N_{V_i}$ of a node. Secondly, they help determine the link reliability of each node with the members of its $N_{V_i}$. In the first phase of clustering, i.e. CH election, each node determines its Mean in-Link Reliability ($MLR_{in}$) which is based upon the initially exchanged NDMs. However, upon reclustering the network, the nodes utilize the messages exchanged during the normal operation of the network to compute both in and out-link reliabilities. This takes into account the time varying nature of the link reliability values for the nodes, since the information used to compute them is gathered over a longer period of time.

---

**Algorithm 3** IDEG (ROLE OF IDEAL DEGREE IN ACDMCP)

1: **if** node1.degree < IDEG ∧ node2.degree < IDEG **then**
2:     prefer the node with a higher degree in becoming a CH
3: **else if** node1.degree > IDEG ∧ node2.degree > IDEG **then**
4:     prefer the node with a lower degree in becoming a CH
5: **else if** node1.degree < IDEG ∧ node2.degree > IDEG **then**
6:     prefer the node with a lower degree in becoming a CH

---

This also adds to the adaptive nature of ACDMCP, since between two clustering periods each node collects statistics on its successful or otherwise message transmissions with its 1-hop neighbors. These statistics are shared within the 1-hop neighborhood in each new round of clustering, so that they could be used to recompute link reliabilities. Thus each node has more reliable data on its communication links which enables it to make better decisions in each successive round of clustering.

One of the design parameters of ACDMCP, which has been included to ensure that only high energy nodes compete for becoming CHs, is the threshold energy ($E_{th}$). This is the bare minimum residual energy of a node which allows it to assume the role of a CH. Since we want well balanced clusters that don't vary in size greatly and which could also minimize the radio signal interference (should we choose some TDMA based MAC scheme inside clusters), we could achieve that by assigning an appropriate value to the IDEG design parameter. This parameter defines our preferred cluster size. In our experiments, we assign it a value of 4 which simply means that the nodes having a degree of 4 are favored in becoming CHs. This value could be changed easily, if the application requirements are to have clusters of some specific size. The way ACDMCP utilizes it in the CH election phase is shown in the Algorithm 3. It ensures that a node, having a degree closer to IDEG, is preferred in assuming the role of CH in its neighborhood.

**Algorithm 4** CLUSTERING (PHASE-1: CH ELECTION)

```
1:  Variables: NDM = Neighbor Discovery Message,
    ICH = I AM CH msg, SNCH = Single Node CH
2:  randomly broadcast NDM n times
3:  wait, for specific time, to receive NDMs
4:  if receivedMsg = NDM then
5:      add sender to your neighborlist and count
        NDMs
6:  if my neighborhood set = φ then
7:      set myStatus = SNCH
8:  else
9:      if reclustering then
10:         share the number of msgs received
            from & sent to each neighbor before
            reclustering
11:     compute MLR_in
12:     if my E_re < E_th then
13:         myStatus = UC
14:     else
15:         myStatus = CHC
16:         share MLR_in, E_re, (D_Vi) within 1-hop
            Neighborhood
17:         compute own and neighbor's CHCV_MLR_in
18:         if my CHCV_MLR_in > all neighbor's
            CHCV_MLR_in ∧ myStatus = CHC then
19:             set myStatus = CH and broadcast
                ICH msg
20:         else if my CHCV_MLR_in = any neighbor's
            CHCV_MLR_in ∧ myStatus = CHC then
21:             compare your MLR_in, E_re, (D_Vi),
                NID in the same order with that of
                your 1-hop neighbors
22:             if you are best, set myStatus = CH
                and broadcast ICH msg
23:         else
24:             set myStatus = UC and wait for ICH
                msg
```

The three parameters that make up the metric $CHCV_{MLR_{in}}$, which is used in CH election phase of the clustering process, are shared with the nodes in the $N_{V_i}$. Each node in the CHC state determines if it is the best suited node to assume the CH role by comparing its $CHCV_{MLR_{in}}$ value with that of its neighbors. Ties are broken by comparing the constituent parameters of the $CHCV_{MLR_{in}}$ metric in the desired order. Currently, we resolve ties by comparing the nodes' $MLR_{in}$, $E_{re}$, $D_{V_i}$, and NID respectively. This order is based on the observation that a node having relatively higher energy can dissipate it quickly if it has poor link reliability with its neighbors in its $N_{V_i}$. Finally, the best node assumes the CH role and broadcasts its cluster joining offer to its 1-hop neighbors to form clusters.

## 5.4. Cluster Formation

In the cluster formation phase, the nodes wait, a specific amount of time, for the CH announcements (cluster joining offers). Upon hearing these offers, each node selects the best offer using the $CHCV_{LR_{out}}$ metric. In this metric out-Link Reliability is used, since it is more relevant in a clustering hierarchy as most of the communication takes place from the nodes to their respective CH. It, however, could be replaced with a metric like Estimated Transmission count (ETX) [14] that includes both in and out link reliabilities, should the application requirements necessitate so. If two offers have the same value for $CHCV_{LR_{out}}$, nodes use out-link reliability, degree, $E_{re}$, NID respectively to break the tie.

Figure 1: Cluster formation process (left) and State-transition diagram of ACDMCP (right).

The nodes that lie outside 1-hop range of the elected CHs cannot hear any CH announcements directly. They, however, receive offers from existing CMs or TCMs to join a cluster transitively. Once a node hears such offers, it uses $CHCV_{ELR_{out}}$ to evaluate them. Here the notion of end-to-end link reliability of the complete d-hop path to the CH is used instead of a greedy approach whereby out-link reliability to the nodes offering transitive cluster membership is used. We choose this end-to-end approach because of its obvious advantage over the greedy approach. The greedy approach would suffer if the multi-hop communication link, up to the CH, has low link reliability after the immediate neighbor. In case of a tie between two offers, nodes use $ELR_{out}$ between themselves and their CH, hop-distance to the CH, degree, $E_{re}$, and NID to break the tie.

The Figure 1 (left) shows in/out-link reliabilities on the edges between nodes. The ELRs are shown with lines that span more than one edge length. The node 18 chooses CH-3 instead of CH-2, since it has a higher out-LR with the former (assuming out-LR is the parameter used for breaking the tie), despite being physically closer to CH-2. Node 17 doesn't have a CH in its 1-hop communication range and it receives transitive membership offers from three of its neighbors which are already CMs of different clusters. Again, we can assume that there is a tie, on the basis of $CHCV_{ELR_{out}}$ metric, which we are resolving using the $ELR_{out}$. If it were to take a greedy approach, it should accept that offer which hast the highest out-LR. On the contrary, it takes an end-to-end approach and compares the offers on ELR basis. The ELRs on two paths have the same value of 0.35. In order to break this tie, it successively compares the other parameters that make up the $CHCV_{ELR_{out}}$ metric, ultimately breaking the tie using $E_{re}$, for instance.

---

**Algorithm 5** CLUSTERING (PHASE-2: CLUSTER FORMATION)

1: Variables: CJN = Cluster Joining Notification(with the chosen CH id), SCJ = Searching Cluster to Join msg, TCMO = Transitive Cluster Membership Offer(unicast), SN = subneighbor, TCJN = Transitive Cluster Joining Notification(with the chosen TCH id)
2: **if** receivedMsg = ICH **then**
3:   **if** myStatus != CH **then**
4:     wait, for specific time, to receive more ICH announcements
5:     evaluate cluster membership offers using $CHCV_{LR_{out}}$, LRO, $(D_{V_i})$, $E_{re}$, NID, in the same order
6:   **else if** myStatus = CH **then**
7:     resolve conflict using CH Election criteria in Algorithm 4
8:   after accepting membership offer, set myStatus to CM and broadcast CJN
9: **if** myStatus = UC ∧ no ICH msg received **then**
10:   broadcast SCJ
11: **if** receivedMsg = SCJ **then**
12:   **if** myStatus = CH **then**
13:     unicast ICH
14:   **else if** myStatus = CM ∨ TCM **then**
15:     unicast TCMO
16: **if** receivedMsg = TCMO ∧ myStatus = UC **then**
17:   wait, for specific time, to receive more TCMOs
18:   evaluate TCMOs using $CHCV_{ELR_{out}}$, $ELR_{out}$, ς, $(D_{V_i})$, $E_{re}$, NID, in the same order
19:   after accepting best TCMO, set myStatus to TCM and broadcast TCJN
20: **if** receivedMsg = CJN **then**
21:   **if** myStatus = CH **then**
22:     **if** CJN.CHid = myid **then**
23:       mark sender as your CM
24:     **else if** sender is already my CM ∧ CJN.CHid != myid **then**
25:       unmark sender as your CM
26:   **else if** (myStatus = CM ∨ TCM) ∧ sender = my SN **then**
27:     unmark sender as your SN
28:   **else if** (myStatus = CM ∨ TCM) ∧ senderid = myCH **then**
29:     set myStatus = UC and broadcast SCJ
30: **if** receivedMsg = TCJN **then**
31:   **if** myStatus = CM ∨ TCM **then**
32:     **if** TCJN.myTCH = myid **then**
33:       mark sender as SN
34:     **else if** TCJN.myTCH != myid ∧ sender is my SN **then**
35:       unmark sender as SN

---

The transitive CHs share the cluster management load, as they register the TCMs with themselves and bear all the responsibility of aggregating and forwarding their data to the CHs. Therefore, the CHs receive one aggregated message from each of their TCHs which represent data of the TCHs and their sub-neighbor(s) both. In case the CH loses its status, the TCMs are informed of this change by their TCHs. The adaptive nature of ACDMCP allows nodes to switch clusters, if they receive a better offer, even after they have accepted an earlier cluster joining offer. This, however, raises some concern regarding the network state consistency, as multiple CHs or TCHs can have a node listed as their CM or TCM respectively. This issue is resolved by exploiting the broadcast nature of the wireless communication as explained below.

**5.4.1. Maintaining Network State Consistency**

We make use of the broadcast nature of the wireless communication to our advantage in ACDMCP and exploit it to ensure network state consistency. For instance, after accepting a cluster joining offer, the node broadcasts either a Transitive Cluster Joining Notification (TCJN) or a Cluster Joining Notification (CJN), depending upon the type of cluster membership (direct or transitive), which includes the NID of the chosen CH/TCH. When a node receives CJN/TCJN, if it finds that it has been selected as a CH/TCH, it marks the sender as its CM or a sub-neighbor (TCM) in case of a transitive membership. If some other CH/TCH had the sender as its CM or TCM before, it unmarks the sender after finding out that it has joined some other cluster as a direct or transitive member. This approach exploits the broadcast nature of wireless communication and has a higher probability of keeping the network in a consistent state than the one where such notifications are unicast.

## 5.5. Inter-cluster Communication

For the sake of Inter-cluster Communication (ICCOM), the CHs form a multi-hop communication overlay in which communication takes place at high power.

**Algorithm 6** CLUSTERING (PHASE-3: INTER-CLUSTER COMMUNICATION)

1: Variables: SCHICC = Searching CH for Inter-Cluster Communication at High Power (HP), FCHICC = Forwarded SCHICC msg at HP, BACKFSHP = Block ACK From Sink at HP(with msg count of FCHICC that sink overheard (IACK)), DSN = Downstream Neighbor, SLPNICC = Searching Low Power Neighbor(LPN) for Inter-Cluster Communication, FLPNICC = Forwarded SLPNICC msg at LP, BACKFSLP = Block ACK From Sink at LP(with msg count of FLPNICC that sink overheard (IACK)), USN = Upstream Neighbor, HTS = hops to sink
2: **if** myStatus = CH **then**
3:    **if** receivedMsg = SCHICC **then**
4:       count received msgs, mark sender(Sink) as your HP DSN, set HTS = 1, wait, for specific time, to receive more SCHICC msgs from Sink
5:       compute in-LR with sink, broadcast FCHICC n times
6:    **if** receivedMsg = BACKFSHP **then**
7:       compute out-LR with sink
8:    **if** receivedMsg = FCHICC ∧ sender's DSN != myid **then**
9:       count received msgs, note its HTS, and wait, for specific time, to receive more FCHICC msgs
10:       compute in-LR with the sender
11:       evaluate IC Comm offers using $CHCV_{ELR_{out/in}}, ELR_{out/in}, \varsigma, E_{re}$, NID
12:       mark selected CH as your HP DSN, set HTS = selected CH's HTS + 1
13:       broadcast FCHICC further n times
14:    **else if** receivedMsg = FCHICC ∧ sender's DSN = myid **then**
15:       mark sender as your USN, after overhearing all FCHICC msgs compute in-LR with sender
16: **if** receivedMsg = SLPNICC **then**
17:    **if** senderid = sinkid **then**
18:       count received msgs, mark sink as your LP DSN, set HTS = 1, send ack to sink
19:       wait, for a defined time, to hear more SLPNICC from the sink
20:       compute in-LR with the sink
21: **if** receivedMsg = BACKFSLP **then**
22:    compute out-LR with the sink
23:    broadcast FLPNICC
24: **if** receivedMsg = FLPNICC ∧ sender's DSN != myid **then**
25:    **if** myStatus = CM ∨ TCM **then**
26:       wait, for a defined time, to hear more FLPNICC from other neighbors
27:       evaluate IC Comm offers
28:       mark selected neighbor as your LP DSN, set HTS = selected neighbor's HTS + 1
29:    **else if** myStatus = CH **then**
30:       evaluate LP IC Comm offers
31:       mark selected neighbor as your LP DSN
32: **else if** receivedMsg = FLPNICC ∧ sender's DSN = myid **then**
33:    mark sender as your USN

The process is started at the sink node, which broadcasts messages at high power that are meant for discovering CHs in its 1-hop range. The sink node broadcasts multiple messages to assess the link quality with the CHs in its 1-hop range. Upon hearing these messages, the CHs discover that they are at 1-hop from the sink node. They further broadcast this message at high power with their hop count from the sink and the address of their downstream neighbor (which is the sink node itself for the CHs that are at 1-hop from it) along with number of times heard from the downstream neighbor, end-to-end in-link reliability with the sink, $E_{re}$ and direct membership count (number of direct CMs). The downstream neighbor overhears this forwarded message (kind of an implicit *ACK* (IACK)) and uses information contained in it to assess its in and out-link reliability with the sender. The sink node and other downstream CHs send a *Block ACK* (BACK) to inform their upstream CHs of the number of times heard from them, information which is useful for them to compute their out-link reliability with their downstream CHs.

Since it is very likely that the sink finds no CH in its 1-hop range, it broadcasts messages at low power also to discover ordinary nodes that lie in its low power 1-hop range. The same procedure is followed by the low power 1-hop neighbors of the sink except for that they send an *ACK* for each ICCOM discovery message that they hear from the sink. This *ACK* is used by the sink to assess its in-link reliability with these nodes. If a CH hears both high power as well as low power ICCOM discovery message sent by the sink, it uses low power to communicate with the later instead of using high power. The sink also sends a BACK at low power to inform its 1-hop neighbors of the number of times it heard from each of them, information that is useful for these nodes to compute their out-link reliability with the sink.

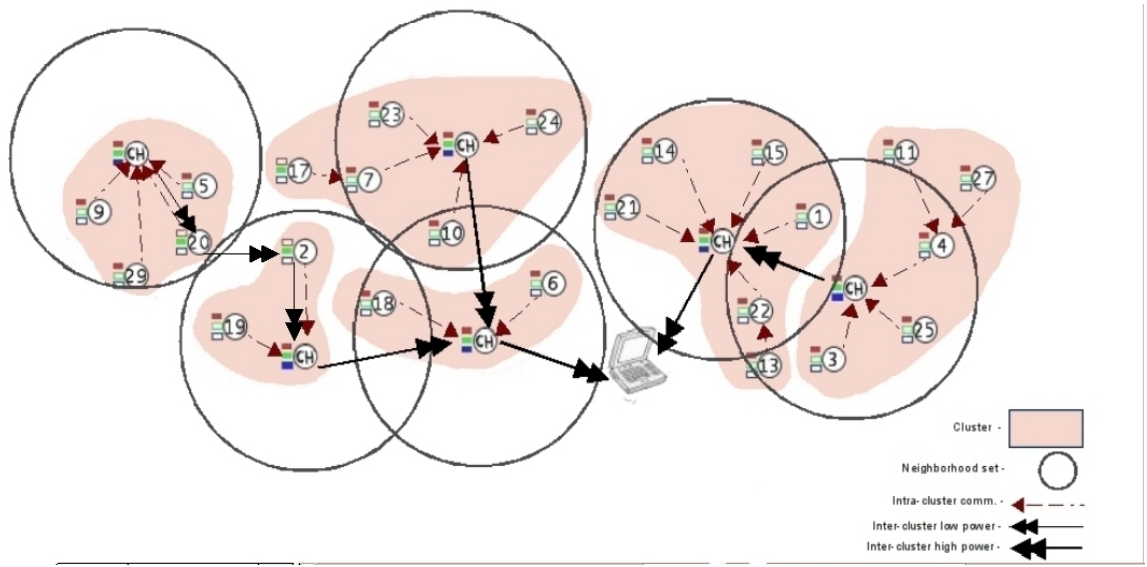

Figure 2: A clustered WSN generated by ACDMCP showing $N_{V_i}$, CHs, direct CMs, TCMs, TCHs, clusters, low and high power communication.

These nodes forward the ICCOM discovery message further so that their CHs can discover routes back to the sink through them. In this way, these nodes serve as gateway nodes for ICCOM when either two CHs can't directly communicate at high power or a CH can't directly reach the sink node at high power. During the normal operation of the network when CHs receive data from their CMs and aggregate it, if needed by the application, and relay it towards the sink using the ICCOM. If a CH finds no high power CH or the sink node in its 1-hop range, it uses one of these gateway nodes and forwards messages at low power to it. The CHs that lie at 1-hop low power range of the sink node also send the data to the sink at low power rather than sending it at high power. This adaptive power control contributes to conserve their battery.

## 6. EVALUATION

We implemented ACDMCP in Contiki [15] which is an open source operating system for programming low power WSN and other embedded systems. Its programming model consists of multiple processes running over an event driven kernel. It also supports multi-threading in the programs through an abstraction called protothreads. Other features include dynamically loadable modules, support for TCP/IP stack, Rime communication stack that offers protocol independent radio networking and a cross-layer network simulation tool called Cooja.

We measure performance of ACDMCP and compare it with a general ID based clustering protocol. The ID based clustering protocol chooses a node as a CH if it is the highest ID node in its neighborhood. The elected CHs advertise their clusters to their 1-hop neighbors to form clusters. The nodes that don't hear any CH announcements choose the highest ID node amongst themselves as their CH.

### 6.1. Simulation Model

We use Cooja to simulate a WSN random deployment without any particular restrictions on node density or distribution. The simulated nodes are tmote sky. The link reliabilities are controlled programmatically to assess the clustering process and inter-cluster routing process of ACDMCP. We simulate different network sizes to see the behavior of the protocol when network scales. The energy consumed by the nodes is assessed using the power profiling mechanism [16] provided by Contiki. We assign the same initial energy to each node in our experiments, unless otherwise stated. The Contiki energy profiling framework measures times for which different components of the nodes remain active. This information, along with current consumption from tmote sky data-sheet, is used to compute the energy consumption by the nodes.

### 6.2. Data Transport Success Ratio

One of our primary objectives for incorporating link reliability in ACDMCP is to ensure the selection of communication paths that could transport data reliably to the sink. For this purpose different incarnations of link reliability have been included throughout the clustering process. In order to measure the performance of ACDMCP on reliability count, we use a simple metric called Data Transport Success Ratio (DTSR). As is evident from the name, DTSR is the ratio of the number of messages that are successfully received at the sink to the total number of messages that are generated by the network. We measure the DTSR achieved by ACDMCP for different network sizes and for different assignments of the IFs in the *CHCV* metric. The results are plotted in Figure 3 (upper-left). The different combinations of IFs (in percent) used are also shown in the legends part of the graph. They appear in the order $IF_{REI}$, $IF_{MLR_{in}}$ and $IF_{NDI}$. It is evident from this plot that ACDMCP does achieve high DTSR when appropriately high IF is assigned to the link reliability parameter in the *CHCV* metric. Additionally, the DTSR is not adversely affected even when the network scales. The slight downward trend that one observes in the graph is due to different network dynamics in each deployment. It should be noted here that no explicit retransmissions are used and each node just sends each message once to its CH or TCH. The DTSRs achieved by the ID-based clustering protocol are much lower than ACDMCP in all network sizes. This is because of the fact that the ID-based clustering protocol remains completely agnostic to the link quality in the clustering process and any good or bad communication paths chosen by it are purely incidental.

### 6.3. Network Lifetime

We measure the performance of the network by counting the number of messages that are received at the sink as well as in milliseconds before major of the nodes die making the network disconnected. To this end, we assume an application that requires nodes to send periodic reports to their respective CH ultimately to be delivered to the sink node. We assume a period of 20 seconds for these reports. We assign all nodes equal initial energy of 10 J in both cases. The performance of both ACDMCP and ID-based clustering protocol is plotted and shown in Figure 3 (lower-part). The plot shows that the performance of ACDMCP is better than ID-based clustering protocol for all network sizes. This difference becomes larger as the network scales because the number of possibilities for choosing bad paths increases for ID-based protocol. The network traffic also grows, thus causing higher interference on unreliable paths resulting in higher losses. The higher number of messages delivered by ACDMCP (Figure 3 lower-right) reflects the fact that it makes better use of the energy resources of the nodes and prolongs the life of the network (also reflected in Figure 3 lower-left). If the same data transport efficiencies are demanded of the ID-based clustering, it would involve higher number of message retransmissions thus impacting the network lifetime adversely.

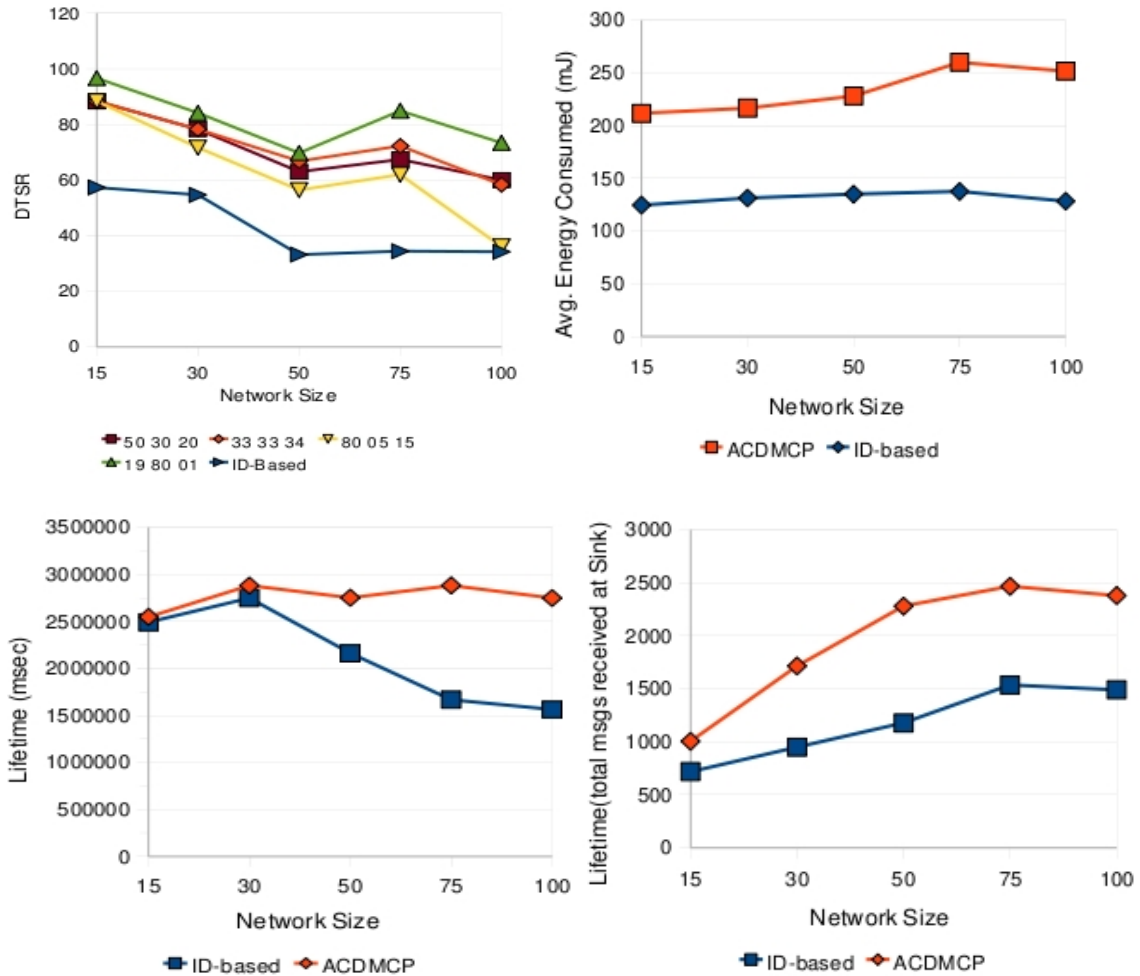

Figure 3: Data transport reliability achieved (upper-left) and energy consumed in a single round of clustering (upper -right). Lifetime of the network measured in milliseconds (lower-left) and total number of messages received at the sink before major of the nodes die (lower-right).

## 6.4. ACDMCP Overhead

The clustering overhead of ACDMCP is also minimal. Since nodes cluster by having local interactions in their 1-hop neighborhood, the scale of the network doesn't impact the clustering overhead in ACDMCP either. The plot in Figure 3 (upper-right) reflects this fact. The energy consumed in the clustering process remains between 210 and 260 mJ showing very small variation with changing size of the network. Please note that this energy overhead is measured when the nodes start the clustering process for the first time. The subsequent rounds of clustering involve even lower overhead, since it is only in the first round of clustering that the nodes exchange explicit NDMs for determining their neighborhood and link reliability in it. However, ACDMCP has a higher energy overhead than ID-based clustering protocol for obvious reasons. The information that is exchanged between the nodes to determine *CHCV* values requires large-sized messages than the ones exchanged in ID-based approach. This small energy overhead is, however, duly compensated by the reliability gains achieved by ACDMCP.

## 7. Applications of Clustering in WSN

Applications of clustering in WSN are quite a few ranging from spatio-temporal in-network aggregation to energy efficient hierarchical routing. One interesting application of clustering in WSNs could be to manage the distribution of event notifications in a publish-subscribe like middleware approach [17]. The CHs could serve as the event notification brokers (notification routers in Figure 4) and can manage the distribution of event notifications to the event consumers as and when they receive a published event by an event producing node in a cluster. If both event producer and consumer are part of the same cluster, then for an event broker delivery of the notification is local to the cluster. However, if the event consumers are spread over multiple clusters, then a federated system of brokers would be needed to spread the event notification. Such a federated broker network could be generated using the inter-cluster communication. Considering scarce resources in a WSN, having global knowledge at each CH of all subscriptions is not feasible. Therefore, each CH could just manage subscriptions that belong to its own cluster. If a published event doesn't have a relevant subscription in the same cluster, the CH can broadcast this event notification at high power to its peers in the federated overlay of CHs. Since only a subset of the nodes (CHs) is involved in forwarding such event notifications, the overall communication costs could be kept low.

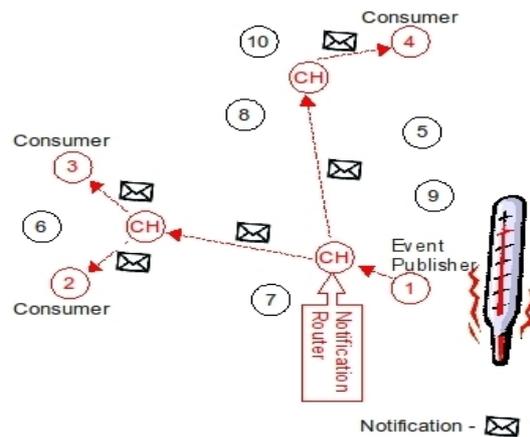

Figure 4: Exploiting clustering hierarchy for the distribution of an event notification in a Publish-Subscribe like middleware approach

Another similar application of clustering is in approaches which model the network as a multiagent system involving both mobile and static agents [1]. In such a system, the mobile agents can be thought of as mobile event producers and consumers and the static agents as the static event producers and consumers that are associated with a particular node in the network. An event detecting agent, event producer in publish-subscribe jargon, that detects an event in a particular cluster can inform the corresponding CH to look for the event consumer agent. If the consumer agent is also present in the same cluster, then event could be communicated to it directly, otherwise the CH can broadcast the event notification to the CHs of neighboring clusters eventually to be delivered to the corresponding event consuming agent.

## 8. Conclusions

We have addressed the problem of developing energy efficient and reliable clustering hierarchies in WSN. The metric used for forming clusters and establishing inter-cluster communication data routing paths incorporates a flexible mix of node residual energy, degree, hop-count and different incarnations of the notion of link reliability. The multi-hop clusters generated by our protocol ensure that the best link reliability communication paths are selected, when the users demand higher data reporting reliability from the network. The adaptive nature of the suggested protocol allows nodes to switch to the best CH within their communication range. It also allows catering for the time varying nature of link reliability by reassessing it using data collected during normal operation of the network.

The clustering overhead is also reasonably small and could easily be ignored considering the reliability benefits achieved by the protocol.

## Acknowledgements

This work has been supported by Higher Education Commission Pakistan, DAAD and TU Darmstadt. The authors wish to thank everybody who helped in whatever way to improve this paper especially Giedre Marozaite (Bachelor Student at the department of Mathematics, TU Darmstadt) who worked hard in bringing the paper in its present shape.